\documentclass[12pt,a4paper]{article}
\usepackage[utf8]{inputenc}
\usepackage[english]{babel}
\usepackage{graphicx}
\usepackage{amsmath,amssymb,amsfonts}
\usepackage{wasysym}
\usepackage{afterpage}
\usepackage{mathpazo}
\usepackage{multicol}
\usepackage{subfigure}
\usepackage{xspace}
\usepackage{colortbl}
\usepackage[hyphens]{url}
\usepackage[margin=1.5ex,bf]{caption}
\usepackage{a4wide}
\usepackage{color}
\usepackage{lineno} 
\usepackage{wrapfig}
\usepackage{cite}
\usepackage{multirow}
\usepackage{upgreek}
\usepackage{siunitx}
\usepackage{hepnames}
\usepackage{enumitem}

\usepackage[normalem]{ulem}
\usepackage{color}
\definecolor{blue}{rgb}{0,0,1}
\definecolor{red}{rgb}{1,0,0}

\definecolor{darkred}{rgb}{0.5,0,0}
\definecolor{darkblue}{rgb}{0,0,0.5}
\definecolor{firebrick}{rgb}{0.75,0.125,0.125}
\definecolor{darkgreen}{rgb}{0,0.5,0}

\newcommand{\NASixtyOne}{NA61\slash SHINE\xspace}

\newcommand{\eV}{\ensuremath{\mbox{e\kern-0.1em V}}\xspace}
\newcommand{\GeV}{\ensuremath{\mbox{Ge\kern-0.1em V}}\xspace}
\newcommand{\MeV}{\ensuremath{\mbox{Me\kern-0.1em V}}\xspace}
\newcommand{\GeVc}{\ensuremath{\mbox{Ge\kern-0.1em V}\!/\!c}\xspace}
\newcommand{\GeVcc}{\ensuremath{\mbox{Ge\kern-0.1em V}\!/\!c^2}\xspace}
\newcommand{\AGeV}{\ensuremath{A\,\mbox{Ge\kern-0.1em V}}\xspace}
\newcommand{\AGeVc}{\ensuremath{A\,\mbox{Ge\kern-0.1em V}\!/\!c}\xspace}
\newcommand{\MeVc}{\ensuremath{\mbox{Me\kern-0.1em V}/c}\xspace}

\newcommand{\CernVM}{\textsc{Cern\-\kern-0.05emVM}\xspace}

\begin{document}

\vspace{-0.2cm}
\begin{center}
{\large\sc European Laboratory for Particle Physics}\\
\vspace{1.7cm}
{\Large Charm Program of NA61/SHINE: \\Motivation and Measurements\footnote{Invited talk presented at the ECT* Workshop on Phase Diagram of Strongly Interacting Matter: From Lattice QCD to Heavy-Ion Collision Experiments, Trento, Italy, November 27, 2017, and Reimei Workshop on Hadronic Resonances and Dense Nuclear Matter, Tokai, Japan, December 11, 2017}}\\
\vspace{1.5cm}
\includegraphics[width=0.2\textwidth]{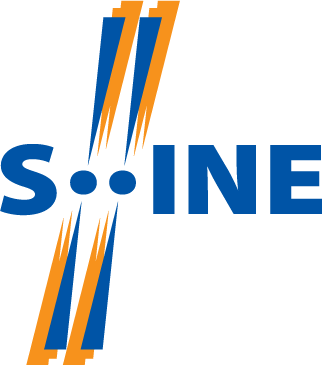}\\
\vspace{1.5cm}
{\large\bf Aleksandra Snoch \\and the NA61/SHINE Collaboration}\\
{\url{http://na61.web.cern.ch/}}
\end{center}

\vspace{1.5cm}

\begin{abstract}
Recently, the \NASixtyOne experiment at the CERN SPS has extended its program on physics of strong interactions by measurements of  charm hadron production in nucleus-nucleus collisions. This charm program is here briefly summarized. 
\end{abstract}
\centerline
\today

\thispagestyle{empty}

\newpage

\begin{center}
	{\Large\textbf{The NA61/SHINE Collaboration}}
\end{center}
\begin{sloppypar}

\noindent
A.~Aduszkiewicz$^{\,16}$,
Y.~Ali$^{\,13}$,
E.V.~Andronov$^{\,22}$,
T.~Anti\'ci\'c$^{\,3}$,
B.~Baatar$^{\,20}$,
M.~Baszczyk$^{\,14}$,
S.~Bhosale$^{\,11}$,
A.~Blondel$^{\,25}$,
M.~Bogomilov$^{\,2}$,
A.~Brandin$^{\,21}$
% ...

\end{sloppypar}
\vspace*{5mm}
{\footnotesize

\noindent
% ...
$^{2}$~Faculty of Physics, University of Sofia, Sofia, Bulgaria\\
$^{3}$~Ru{\dj}er Bo\v{s}kovi\'c Institute, Zagreb, Croatia\\
% ...
$^{11}$~H. Niewodnicza\'nski Institute of Nuclear Physics of the
      Polish Academy of Sciences, Krak\'ow, Poland\\
% ...
$^{13}$~Jagiellonian University, Cracow, Poland\\
$^{14}$~AGH - University of Science and Technology, Cracow, Poland\\
% ...
$^{16}$~University of Warsaw, Warsaw, Poland\\
% ...
$^{20}$~Joint Institute for Nuclear Research, Dubna, Russia\\
$^{21}$~National Research Nuclear University (Moscow Engineering Physics Institute), Moscow, Russia\\
$^{22}$~St. Petersburg State University, St. Petersburg, Russia\\
% ...
$^{25}$~University of Geneva, Geneva, Switzerland\\
% ...
}

\newpage

\section{Introduction}

\NASixtyOne is a fixed-target experiment \cite{Abgrall:2014fa} operating at the CERN Super-Proton-Synchrotron (SPS). The \NASixtyOne Collaboration studies, in particular, properties of hadron production in nucleus-nucleus collisions. The primary aim is to uncover features of the phase transition between confined matter and quark gluon plasma (QGP). Within the current program, data on p+p, Be+Be, Ar+Sc, Xe+La, and Pb+Pb collisions at beam momenta in the range 13\textit{A}-150\AGeVc has been recorded.

This program was recently extended by measurements of  charm hadron production in nucleus-nucleus collisions. The motivation of the charm program is discussed in Sec. \ref{sc:motivation}. The current status of the program including the preliminary results of first data taking campaigns performed in 2016 and 2017 is presented in Sec. \ref{sc:firstmeasurement}. The plans of  NA61/SHINE for systematic charm measurements in the years $ 2021-2024 $ are discussed in Sec. \ref{sc:futureprogram}.

\section{Motivation for the charm program} \label{sc:motivation}

There are a lot of questions that need to be answered concerning charm production in nucleus-nucleus collisions. Among them, there are three that motivate charm measurements by NA61/SHINE:
\begin{enumerate}[label=(\roman*)]
	\item What is the mechanism of charm production? 
	\item How does the onset of deconfinement impact charm production? 
	\item How does the formation of quark gluon plasma impact $ \text{J}/\psi $ production?
\end{enumerate}

To answer these questions, knowledge on the mean number of charm quark pairs produced in full phase space in heavy ion collisions, $ \langle \Pcharm\APcharm \rangle $, is needed. Such data does not exist yet and \NASixtyOne aims to provide them within the coming years.

\subsection{Mechanism of charm production} \label{ssc:charmproduction}

\begin{figure}[!htb]
	\centering
	\includegraphics[width=1\textwidth]{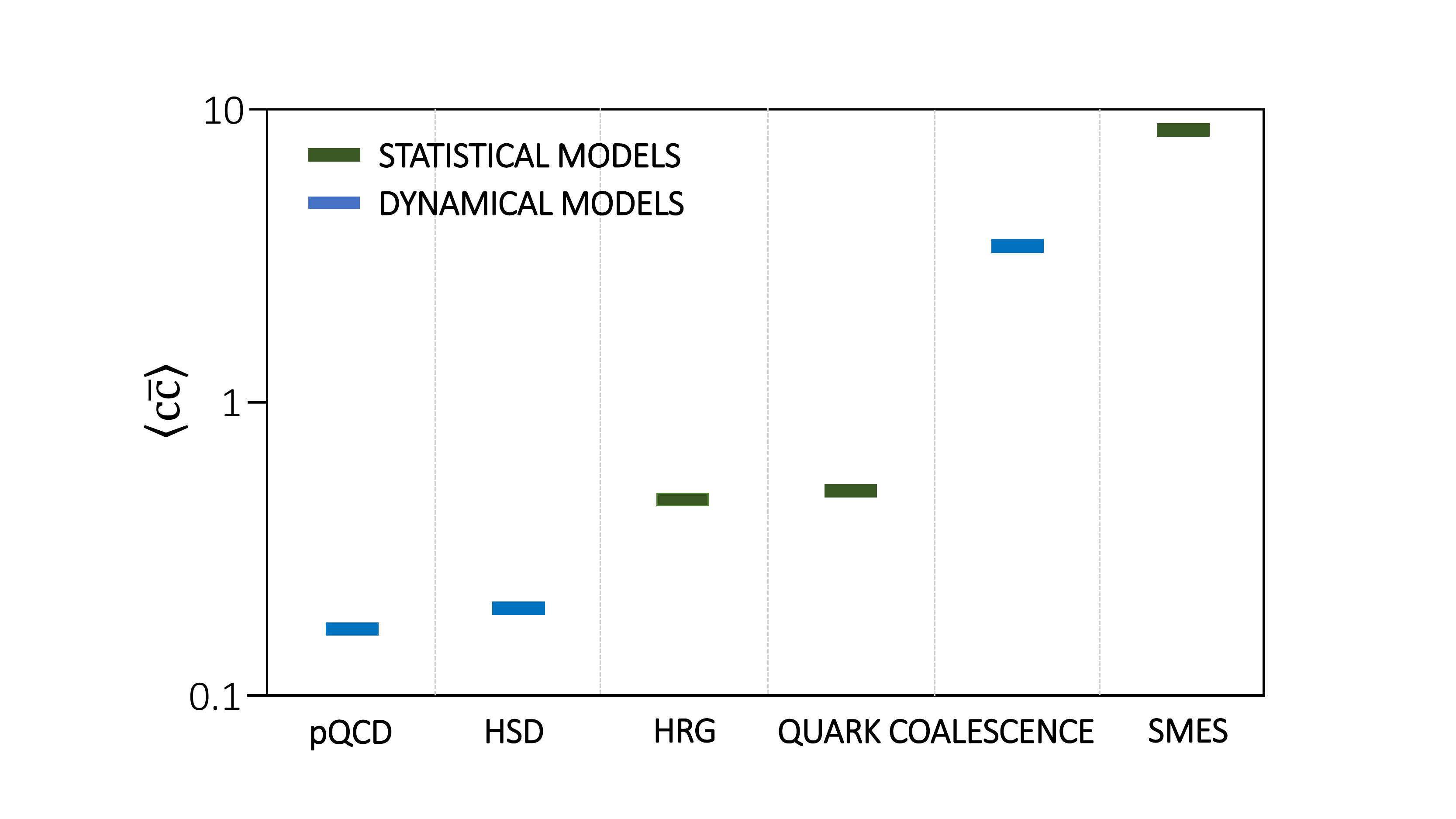}
	\caption{Mean multiplicity of charm quark pairs produced in the full phase space in central Pb+Pb collisions at 158\AGeVc calculated with dynamical models (blue bars): pQCD-inspired \cite{Gavai:1994gb,BraunMunzinger:2000px}, HSD \cite{Linnyk:2008hp}, and Dynamical Quark Coalescence \cite{Levai:2000ne}, as well as statistical models (green bars): HRG \cite{Kostyuk:2001zd}, Statistical Quark Coalescence \cite{Kostyuk:2001zd}, and SMES \cite{Gazdzicki:1998vd}.}
	\label{fig:models}
\end{figure}

Figure \ref{fig:models} presents a compilation of predictions provided by dynamical and statistical models on $ \langle \Pcharm\APcharm \rangle $ in central Pb+Pb collisions at 158\AGeVc. These predictions are obtained from:
\begin{enumerate}[label=(\roman*)]
	\item A pQCD-inspired model \cite{Gavai:1994gb,BraunMunzinger:2000px} - charm data independent calculation based on model assumptions and nucleon pdfs only.
	\item The Hadron String Dynamics (HSD) model \cite{Linnyk:2008hp} - a pQCD-inspired extrapolation of p+p data.
	\item The Hadron Resonance Gas model (HRG) \cite{Kostyuk:2001zd} - a calculation of equilibrium yields of charm hadrons assuming parameters of a hadron resonance gas fitted to mean multiplicities of light hadrons.
	\item The Statistical Quark Coalescence model \cite{Kostyuk:2001zd} - the mean number of $ \langle \Pcharm\APcharm \rangle $ pairs is calculated using the measured $ \langle \text{J}/\psi \rangle $ multiplicity \cite{Abreu:2000ni} and the probability of a single $ \Pcharm\APcharm $ pair hadronization into $ \langle \text{J}/\psi \rangle $ calculated within the model.
	\item The Dynamical Quark Coalescence model \cite{Levai:2000ne} -  the mean number of $ \langle \Pcharm\APcharm \rangle $ pairs is calculated using the measured $ \langle \text{J}/\psi \rangle $ multiplicity \cite{Abreu:2000ni} and the probability of a single $ \Pcharm\APcharm $ pair hadronization into $ \langle \text{J}/\psi \rangle $ calculated within the model.
	\item The Statistical Model of the Early Stage (SMES) \cite{Gazdzicki:1998vd} - the mean number of charm quarks is calculated assuming an equilibrium QGP at the early stage of the collision.
\end{enumerate}

The predictions of the models on $ \langle \Pcharm\APcharm \rangle $ differ by about two orders of magnitude. Therefore, obtaining precise data on $ \langle \Pcharm\APcharm \rangle $ is expected to allow to narrow the spectrum of viable theoretical models and thus learn about the charm quark and hadron production mechanisms. 

\subsection{Charm production as a signal of onset of deconfinement} \label{ssc:signalofdeconfinement}

The production of charm is expected to be different in confined and deconfined matter. This is caused by different properties of charm carriers in these phases. In confined matter the lightest charm carriers are \PD mesons, whereas in deconfined matter the carriers are charm quarks. Production of a $ \PD\APD $ pair ($ 2\mathrm{m_D} = \SI{3.7}{\giga\electronvolt} $) requires energy about \SI{1}{\giga\electronvolt} higher than production of a $ \Pcharm\APcharm $ pair ($ 2\mathrm{m_c} = \SI{2.6}{\giga\electronvolt} $). The effective number of degrees of freedom of charm hadrons and charm quarks is similar \cite{Poberezhnyuk:2017ywa}. Thus, more abundant charm production is expected in deconfined than in confined matter. Consequently, in analogy to strangeness \cite{Gazdzicki:1998vd, Rafelski:1982pu}, a change of collision energy dependence of $ \langle \Pcharm\APcharm \rangle $ may be a signal of onset of deconfinement.

\begin{figure}[!htb]
	\centering
	\includegraphics[width=0.75\textwidth]{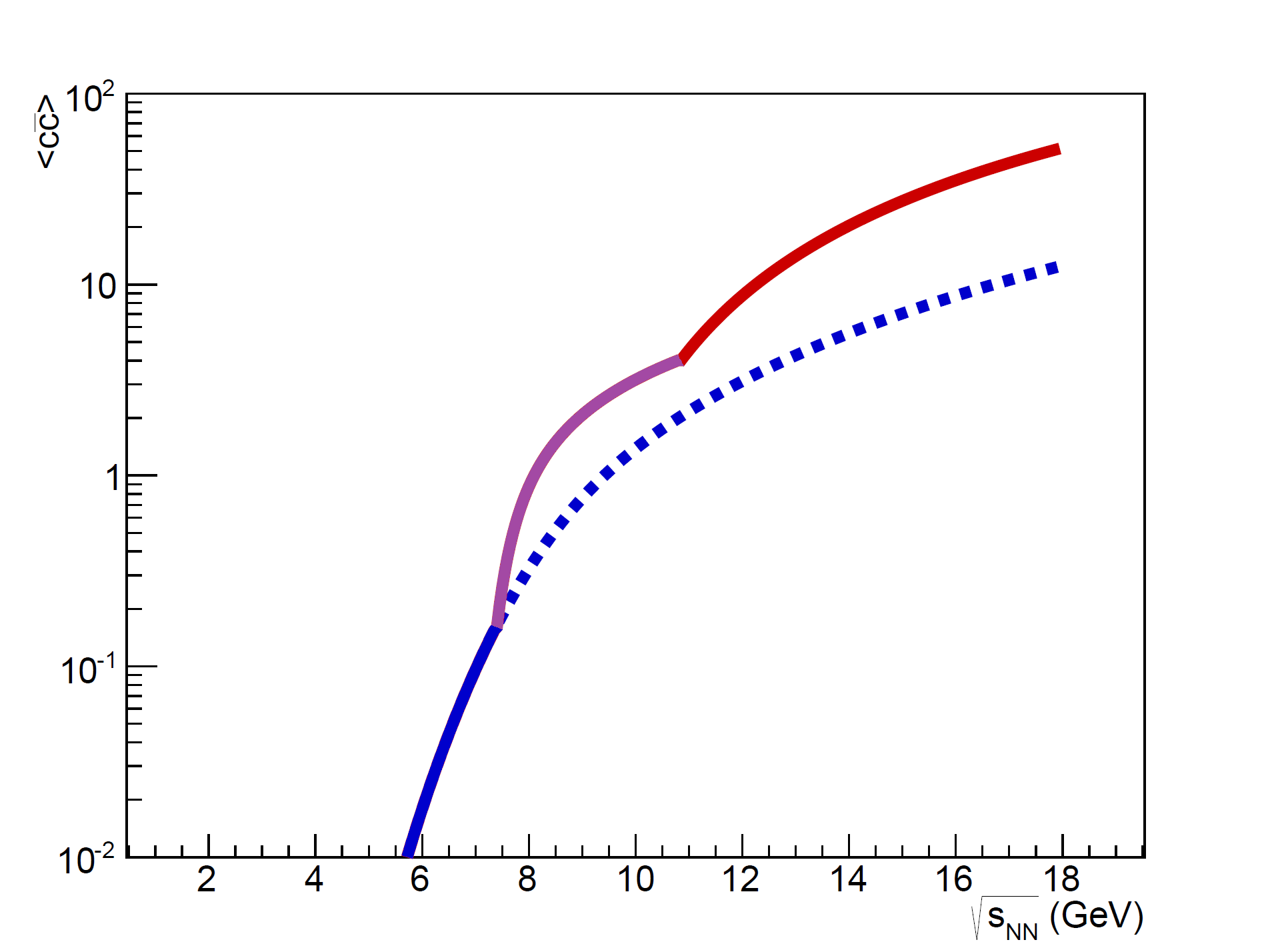}
	\caption{Energy dependence of $ \langle \Pcharm\APcharm \rangle $ in central Pb+Pb collisions calculated within the SMES model \cite{Poberezhnyuk:2017ywa, RomanPrivate}. The blue line corresponds to confined, the purple line to mixed phase, and the red line to deconfined matter. The dashed line presents the prediction without a phase transition.}
	\label{fig:signalQGPsmes}
\end{figure}

\begin{figure}[!htb]
	\centering
	\includegraphics[width=0.6\textwidth]{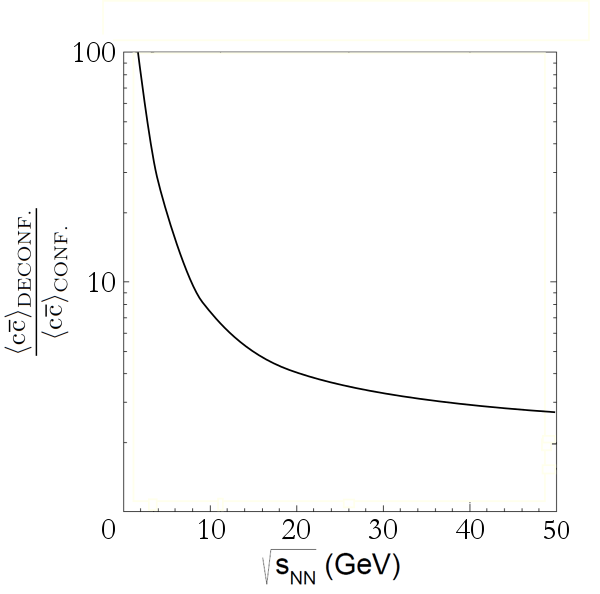}
	\caption{Energy dependence of the ratio of $ \langle \Pcharm\APcharm \rangle $ in deconfined and confined matter in central Pb+Pb collisions calculated within the pQCD-inspired model of Ref.~\cite{Kostyuk:2001sh}.}
	\label{fig:signalQGPpqcd}
\end{figure}

Figures \ref{fig:signalQGPsmes} and \ref{fig:signalQGPpqcd} present collision energy dependence of charm production in central Pb+Pb collisions at 150\AGeVc predicted by two very different models: the Statistical Model of the Early Stage \ref{fig:signalQGPsmes} \cite{Poberezhnyuk:2017ywa} and a pQCD-inspired model \cite{Kostyuk:2001sh}, respectively. 

Figure \ref{fig:signalQGPsmes} shows the energy dependence of $ \langle \Pcharm\APcharm \rangle $ predicted by the Statistical Model of the Early Stage. According to this model, when crossing the phase transition energy range ($ \sqrt{s_{\text{NN}}} = 7-11~\GeV $), an enhancement of $ \langle \Pcharm\APcharm \rangle $ production should be observed. At~150\AGeVc ($ \sqrt{s_{\text{NN}}} = 16.7~\GeV $) an enhancement by a factor of about 4 is expected.

Figure \ref{fig:signalQGPpqcd} shows the ratio of mean multiplicity of $  \Pcharm\APcharm  $ pairs in deconfined and confined matter. Both numerator and denominator were calculated at the same collision energy. At 150\AGeVc ($ \sqrt{s_{\text{NN}}} = 16.7~\GeV $) an enhancement by a factor of about 3 is predicted.

Accurate experimental results will allow to test these predictions.

\subsection{$ \text{J}/\psi $ production as a signal of deconfinement}

Suppresion of the production of $ \text{J}/\psi $ mesons in central Pb+Pb collisions at 158\AGeVc was an important argument for the CERN announcement of the discovery of a new state of matter \cite{Heinz:2000bk}. Within the Matsui-Satz model \cite{Matsui:1986dk} the suppression is supposed to be caused by the formation of the QGP.

\begin{figure}[!htb]
	\centering
	\includegraphics[width=0.9\textwidth]{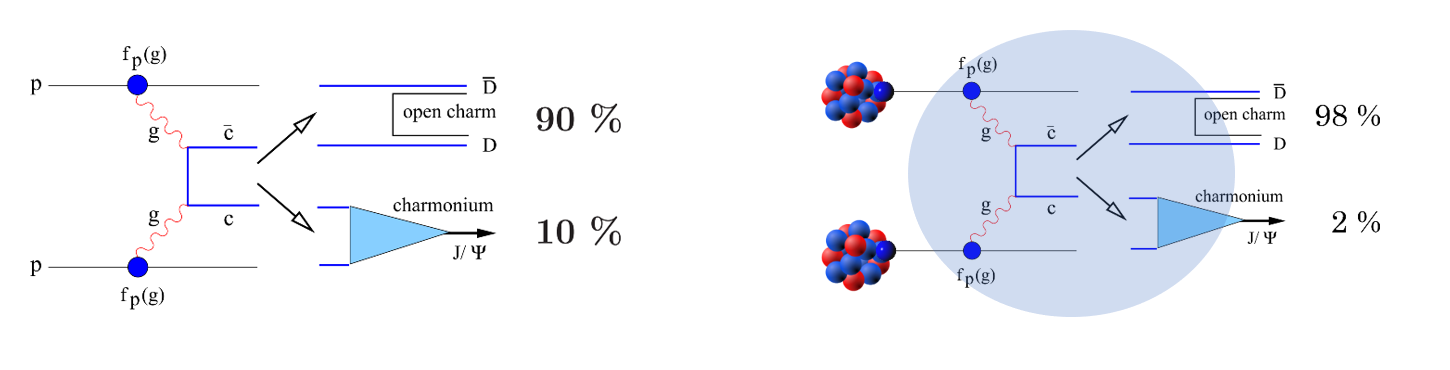}
	\caption{Sketch of the $ \text{J}/\psi $ production mechanism and its relation to $\Pcharm\APcharm$ production in p+p interactions (\textit{left}) and central heavy ion collisions (\textit{right}) \cite{Satz:2013ama}.}
	\label{fig:JpsiProd}
\end{figure}

Figure \ref{fig:JpsiProd} presents two scenarios of charmonium production. In the first case (Fig. \ref{fig:JpsiProd} (\textit{left})), a produced $\Pcharm\APcharm$ pair hadronizes in a vacuum -- this corresponds to a p+p interaction. Open charm and charmonia are produced in vacuum with a certain probability, at high collision energies typically 10\% of $\Pcharm\APcharm$ pairs form charmonia and 90\% appear in charm hadrons.

The second scenario is presented in Fig. \ref{fig:JpsiProd} (\textit{right}). Here the $\Pcharm\APcharm$ pair forms a pre-charmonium state in the quark gluon plasma. Due to the color screening, which may lead to disintegration of this state, the probability of charmonium production is suppresed in favor of open charm production.

The probability of a $\Pcharm\APcharm$ pair hadronizing to $ \text{J}/\psi $ is defined as:
\begin{equation}
P \left(\Pcharm\APcharm \to \text{J}/\psi \right) \equiv \frac{\langle \text{J}/\psi \rangle}{\langle \Pcharm\APcharm \rangle}\equiv\frac{\sigma_{\text{J}/\psi}}{\sigma_{\Pcharm\APcharm}}	.
\end{equation}

To be able to calculate this probability, one needs data on both $ \text{J}/\psi $ and $\Pcharm\APcharm$ yields in the full phase space. At the CERN SPS precise $ \langle \text{J}/\psi \rangle $ data was provided by the NA38~\cite{Abreu:2000nj}, NA50 \cite{Abreu:2000ni}, and NA60 \cite{PhysRevLett.99.132302} experiments, while $ \langle \Pcharm\APcharm \rangle $ data is not available at the CERN SPS up to now.

The problem of the lack of the $ \langle \Pcharm\APcharm \rangle $ data was worked around \cite{Matsui:1986dk, Abreu:2000ni} by assuming that the mean multiplicity of $\Pcharm\APcharm$ pairs is proportional to the mean multiplicity of Drell-Yan pairs: $ \langle \Pcharm\APcharm \rangle \sim \langle DY \rangle $.

\begin{figure}[!htb]
	\centering
	\includegraphics[width=0.45\textwidth]{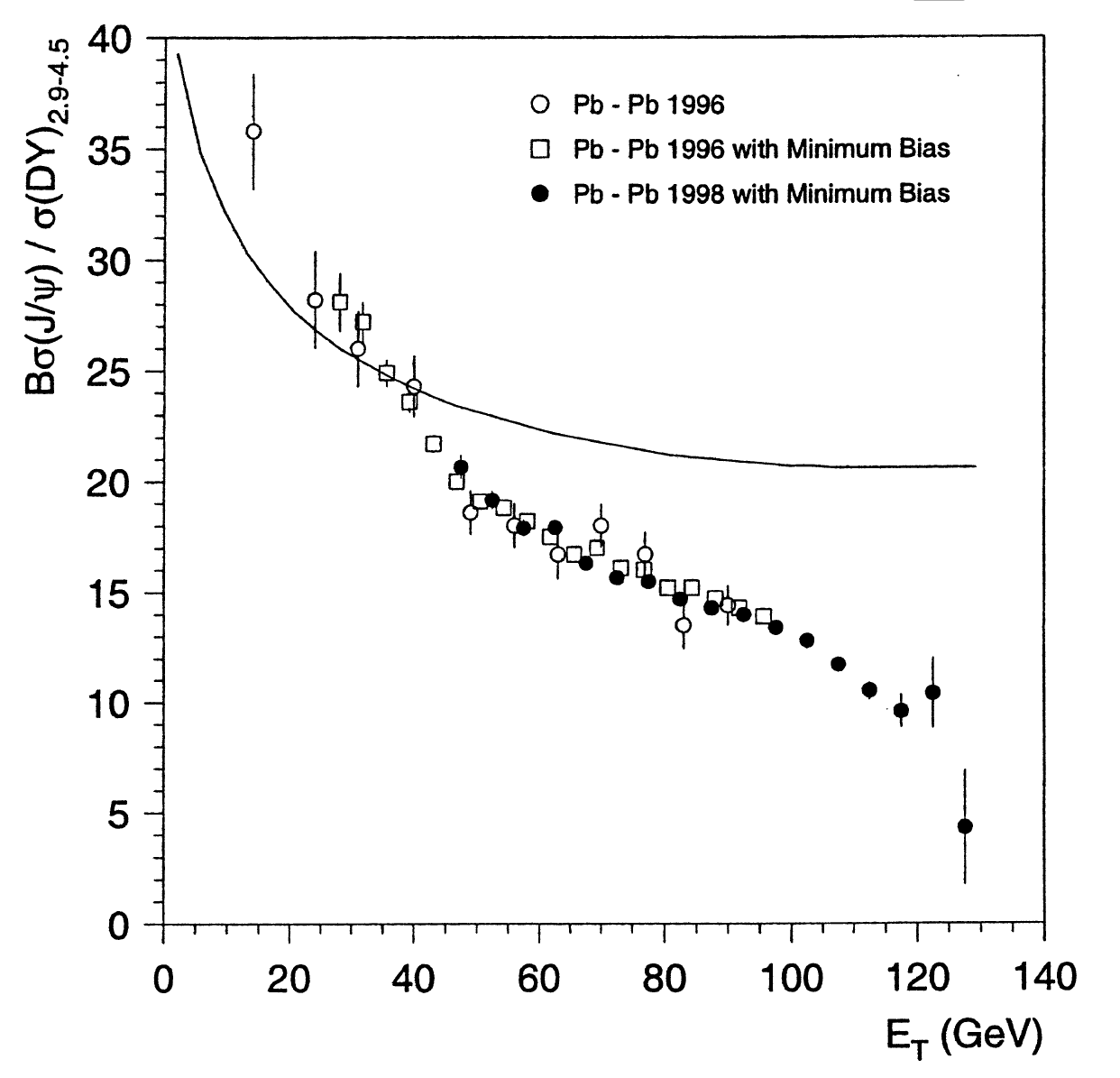}
	\caption{The ratio of $\sigma_{\text{J}/\psi}/\sigma_{\mathrm{DY}}$ as a function of transverse energy in Pb+Pb collisions at 158\AGeV measured by NA50. The curve represents the $ \text{J}/\psi $ suppression due to ordinary nuclear absorption \cite{Abreu:2000ni}.}
	\label{fig:na50Jpsi}
\end{figure}
\newpage
Figure \ref{fig:na50Jpsi} shows the result from the NA50 experiment \cite{Abreu:2000ni} that was interpreted, basing on this assumption, as an evidence for QGP creation in central Pb+Pb collisions at 158\AGeV. However, the assumption $ \langle \Pcharm\APcharm \rangle \sim \langle \text{DY} \rangle $ may be incorrect due to many effects, such as shadowing or parton energy losses \cite{Satz:2014usa}.

This clearly shows a need for precise data on $ \langle \Pcharm\APcharm \rangle $. NA61/SHINE has recently started such measurements.

\section{First measurements of open charm in \NASixtyOne} \label{sc:firstmeasurement}

Precise measurements of charm hadron production by \NASixtyOne should be performed in 2021-2024. The related preparations have started already. In 2015 and 2016, a~Small Acceptance Vertex Detector was constructed and the first measurements of open charm has been started in 2016.

\subsection{Small Acceptance Vertex Detector} \label{ssc:savd}

\begin{figure}[!htb]
	\centering
	\includegraphics[width=0.9\textwidth]{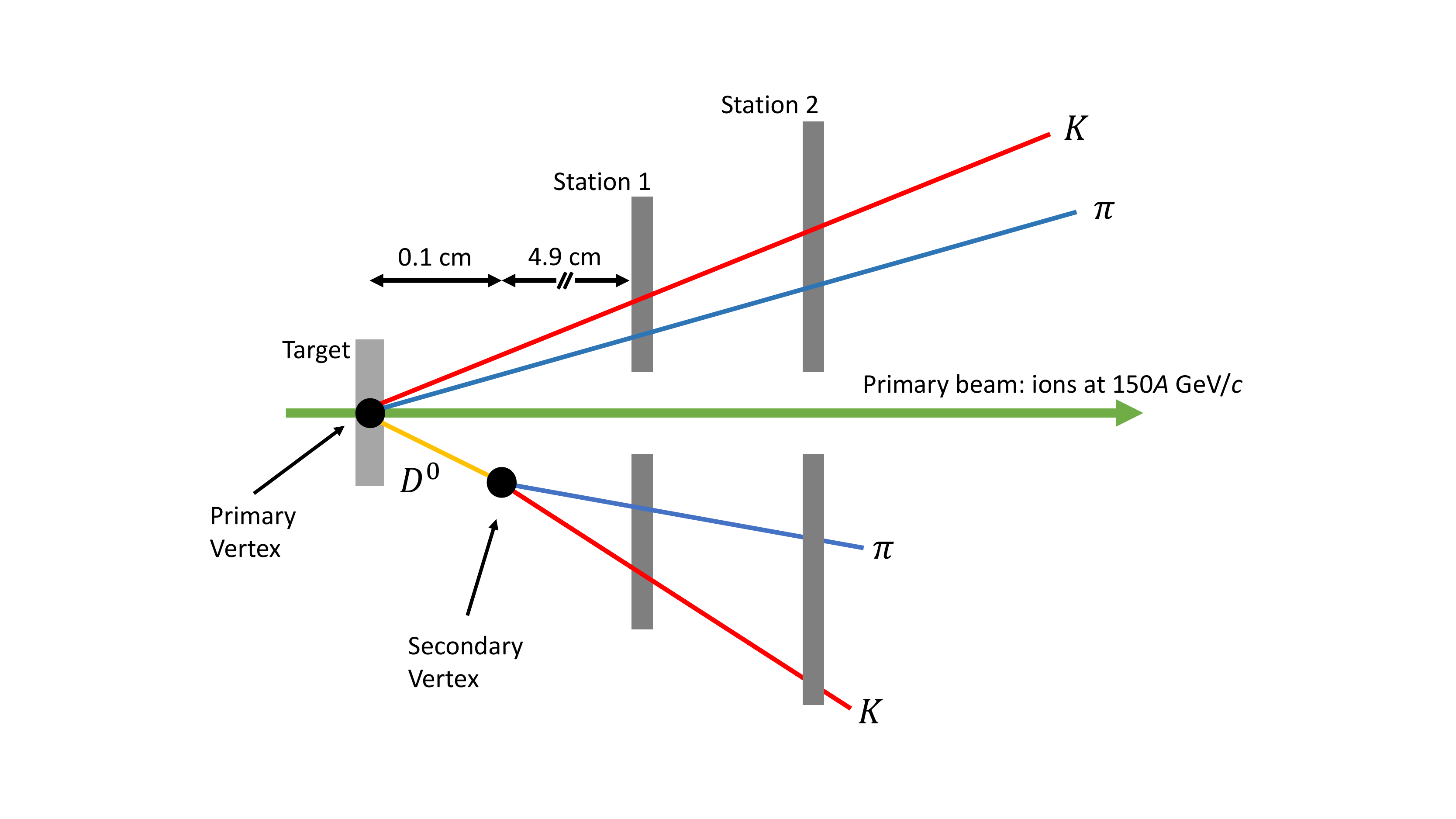}
	\caption{Schematics of reconstruction of a $\PDzero \to \pi^+ + \PKminus$ decay with help of a~vertex detector.}
	\label{fig:VDscheme}
\end{figure}

Starting the open charm program required construction of a new high-resolution vertex detector. Its role in measurements of charm hadrons is schematically shown in Fig. \ref{fig:VDscheme}.

This measurement is challenging due to the short mean lifetime of charm hadrons and the relatively low branching ratios into reconstructable decay channels. Table \ref{tab:opencharm} presents properties of the more frequently produced charm hadrons relevant to their measurement.

\renewcommand{\arraystretch}{2}

\begin{table}[!htb]
	\caption{The most frequently produced charm hadrons: their mass, mean life time, and best suited decay channels for measurements are shown.}
	\label{tab:opencharm}
	\centering
	\begin{tabular}{c l c c}
		Hadron & Decay channel & $c\bar{\tau}$ [\si{\micro\meter}] &  BR \\ \hline
		\PDzero & $\pi^+ + \PKminus$ & 123 &  3.89\% \\
		\PDplus & $\pi^+ + \pi^ + \PKminus $ & 312 &  9.22\% \\
		\PDsplus & $\pi^+ + \PKminus + \PKplus$ & 150 &   5.50\% \\
		$ \Lambda_{\text{c}} $ & $\Pproton + \pi^+ + \PKminus$ & 60 &   5.00\% \\
	\end{tabular}
\end{table}

A big upside of \NASixtyOne is the fact, that this is a fixed-target experiment. Due to the Lorenz boost ($ \beta\gamma \approx 10 $ at midrapidity and $ p_\text{T} \approx 0 $), the average separation between the primary and the decay vertices of \PDzero mesons is about \SI{1}{\milli\meter}. This makes the measurement significantly easier than in the case of collider experiments. In addition, due to the fact that magnetic field is perpendicular to the beam direction (unlike in a typical collider experiment, where it is parallel to the beam direction), the acceptance extends to $p_\text{T} = 0$.

\begin{figure}[!htb]
	\centering
	\includegraphics[width=0.75\textwidth]{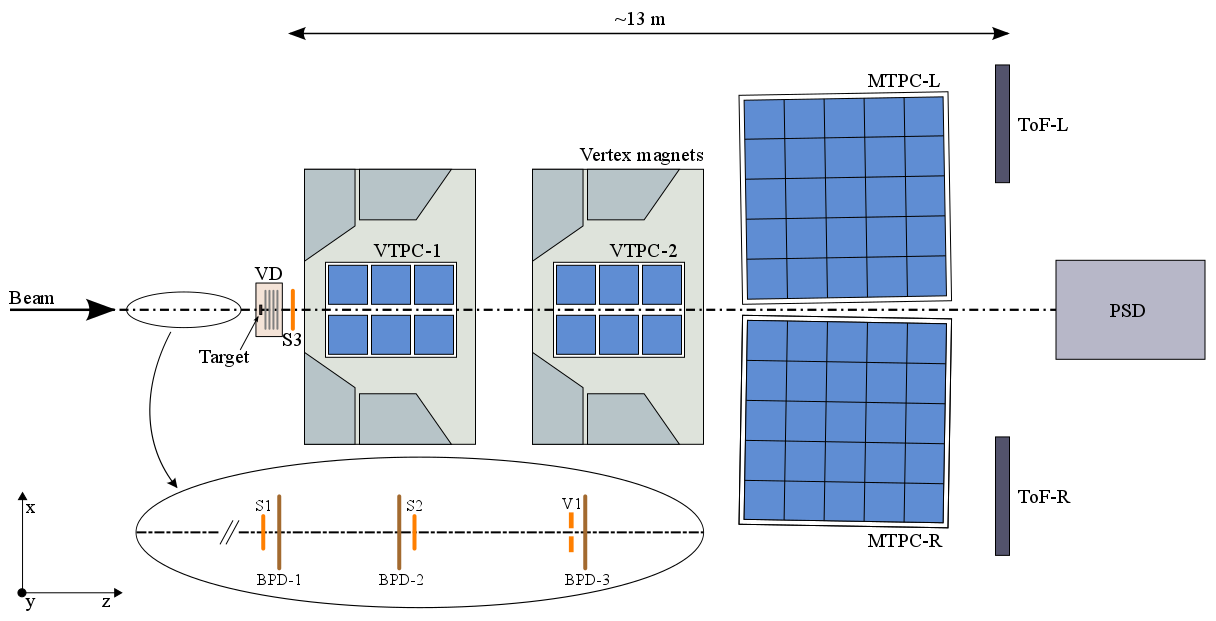}
	\caption{Schematic layout of the NA61/SHINE experiment at the CERN SPS (horizontal cut in the beam plane, not to scale). The beam and trigger counter configuration used for data taking on Pb+Pb collisions in 2016 is presented. The chosen right-handed coordinate system is shown on the plot. The incoming beam direction is along the \textit{z} axis. The Small Acceptance Vertex Detector together with the integrated target station is located upstream of the VTPC-1.}
	\label{fig:VDplacement}
\end{figure}

For the measurement of \PDzero mesons, the Small Acceptance Vertex Detector (SAVD) was added to the \NASixtyOne detector in October 2016. It's location in the \NASixtyOne detector is shown in Fig.~\ref{fig:VDplacement}.

The SAVD was built using sixteen CMOS MIMOSA-26 sensors \cite{mimosa26}. The basic sensor properties are:
\begin{enumerate}[label=(\roman*)]
	\item $ 18.4 \times \SI{18.4}{\micro\meter^2} $ pixels, 
	\item \SI{115}{\micro\second} time resolution,
	\item $ 10 \times \SI{20}{\milli\meter^2} $ surface, 0.66 MPixel,
	\item \SI{50}{\micro\meter} thick.	
\end{enumerate}
The estimated material budget per layer is $ 0.3\% $ of the radiation length.

The sensors were glued to eight ALICE ITS ladders \cite{Abelev:1625842}, which were mounted on two horizontally movable arms and spaced by \SI{5}{\centi\meter} along the \textit{z} (beam) direction. The detector box was filled with He (to reduce beam-gas interactions) and contained an integrated target holder to avoid unwanted material and multiple coulomb scattering between target and detector.

\begin{figure}[!htb]
	\centering
	\includegraphics[width=0.55\textwidth]{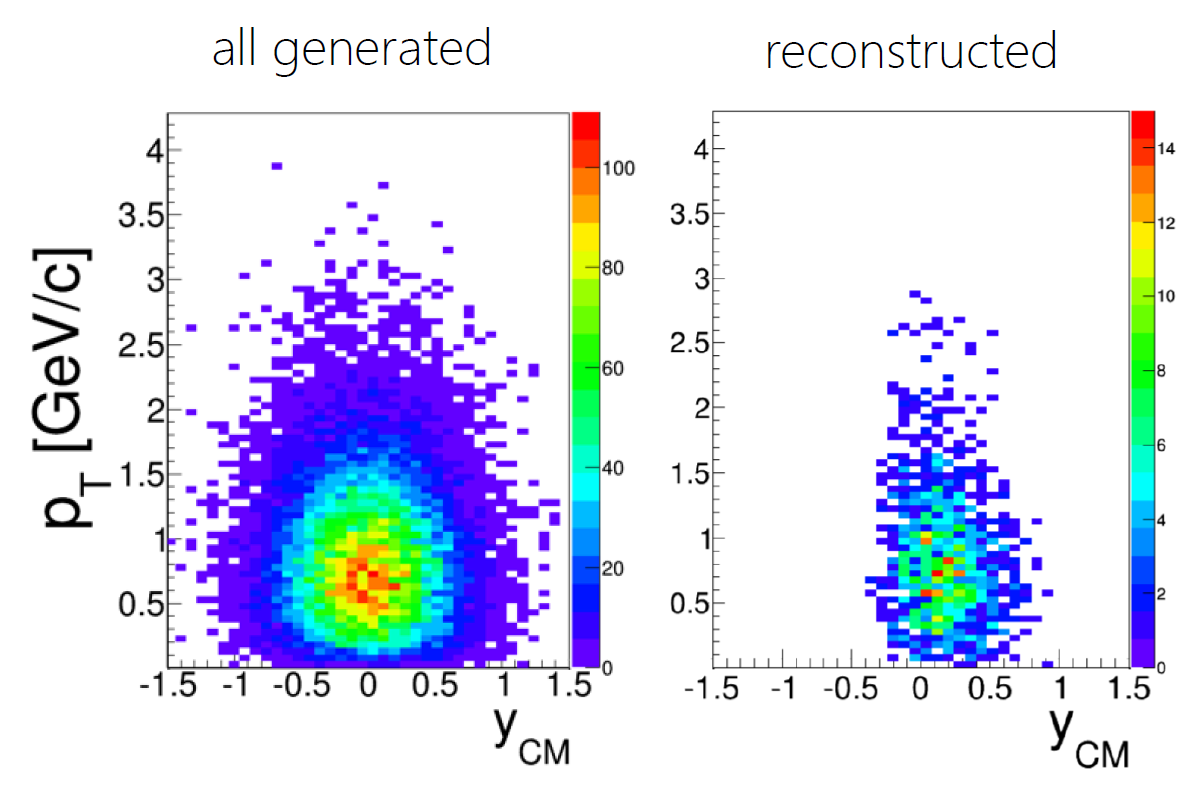}
	\caption{The AMPT model simulation of transverse momentum and rapidity distributions of \PDzero mesons produced central Pb+Pb collisions at 150\AGeVc corresponding to $4\cdot10^6$ events. \textit{Left}: all produced \PDzero mesons. \textit{Right}: \PDzero mesons fulfilling the following criteria: decay into $\PDzero \to \pi^+ + \PKminus$, both decay products registered by the SAVD, and passing background suppression and quality cuts \cite{Aduszkiewicz:2059811}.}
	\label{fig:acceptSim}
\end{figure}

The simulations of SAVD performance using the AMPT physics
input \cite{Lin:2004en} has shown that about 4\% \cite{Aduszkiewicz:2059811} of all $\PDzero \to \pi^+ + \PKminus$ decays will be reconstructed and accepted by the analysis cuts. Figure \ref{fig:acceptSim} shows the phase space coverage which the SAVD provides.

\newpage
\subsection{Test data taking in 2016: Pb+Pb central collisions at 150\AGeVc} \label{ssc:PbPb2016}

The SAVD was used in December 2016 during a Pb+Pb test run. Data on central Pb+Pb collisions at 150\AGeVc were collected. Using these data, the following was demonstrated:
\begin{enumerate}[label=(\roman*)]
	\item tracking in a large track multiplicity environment,
	\item precise primary vertex reconstruction,
	\item TPC and SAVD track matching,
	\item feasibility to search for the \PDzero and \APDzero signals.
\end{enumerate}

\begin{figure}[!htb]
	\centering
	\includegraphics[width=0.58\textwidth]{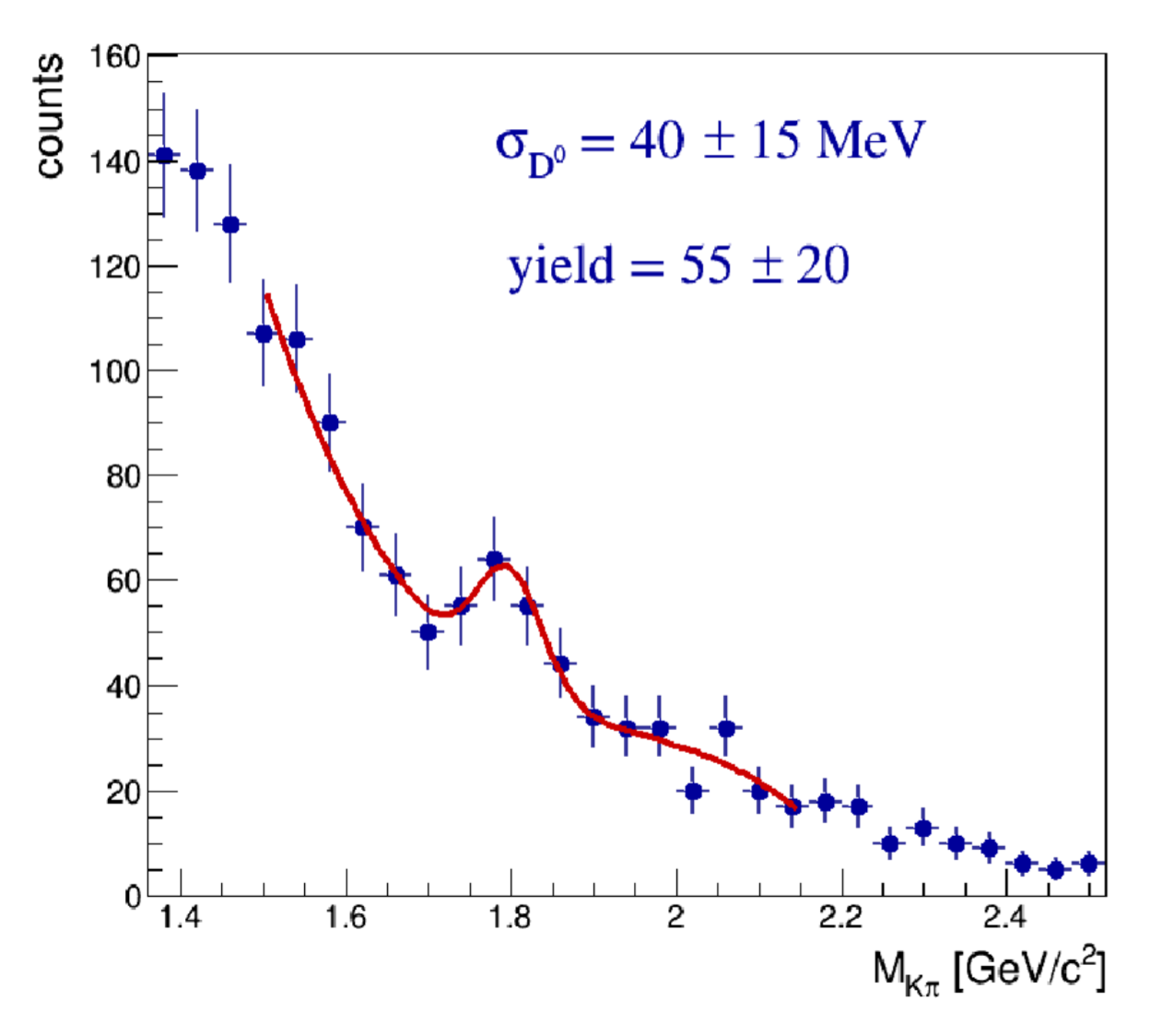}
	\caption{The invariant mass distribution of \PDzero and \APDzero candidates in central Pb+Pb collisions at 150\AGeVc after the background suppression cuts. The particle identification capability of \NASixtyOne was not used at this stage of the analysis \cite{Aduszkiewicz:2059811}.}
	\label{fig:firstD0}
\end{figure}

Based on these data, the spatial resolution of the SAVD was determined. Cluster position resolution is $\sigma_{\text{x,y}}(\text{Cl}) \approx \SI{5}{\micro\metre} $ and primary vertex resolution in the transverse plane is $\sigma_{\text{x}}(\mathrm{PV}) \approx \SI{5}{\micro\metre} $, $\sigma_{\text{y}}(\text{PV}) \approx \SI{1.8}{\micro\metre} $\footnote{$\sigma_{\text{x}}(\text{PV}) > \sigma_{\text{y}}(\text{PV})$ because $B_\text{y} \gg B_\text{x}$}, and along the beam direction is $\sigma_{\text{z}}(\text{PV}) \approx \SI{30}{\micro\metre} $ for a typical multiplicity of events recorded in 2016. Primary vertex resolution of \SI{30}{\micro\metre} is sufficient to perform the search for the \PDzero and \APDzero signals. Figure \ref{fig:firstD0} shows the first indication of a \PDzero and \APDzero peak obtained using the data collected during the Pb+Pb run in 2016.

\subsection{Data taking in 2017 and 2018} \label{ssc:XeLa2017} \label{ssc:PbPb2018}

Successful performance of the SAVD in 2016 led to the decision to also use it during the Xe+La data taking in 2017. About $5\cdot10^6$ events on central Xe+La collisions at 150\AGeVc were collected in October and November 2017.

To get a first estimate of the number of \PDzero and \APDzero decays that can be reconstructed in this data set, simulations for Pb+Pb collisions and p-QCD inspired system size dependence were combined. Based on these simulations, one expects to reconstruct several hundred of \PDzero and \APDzero decays. This would be in itself an important physics result.

\begin{figure}[!htb]
	\centering
	\includegraphics[width=0.7\textwidth]{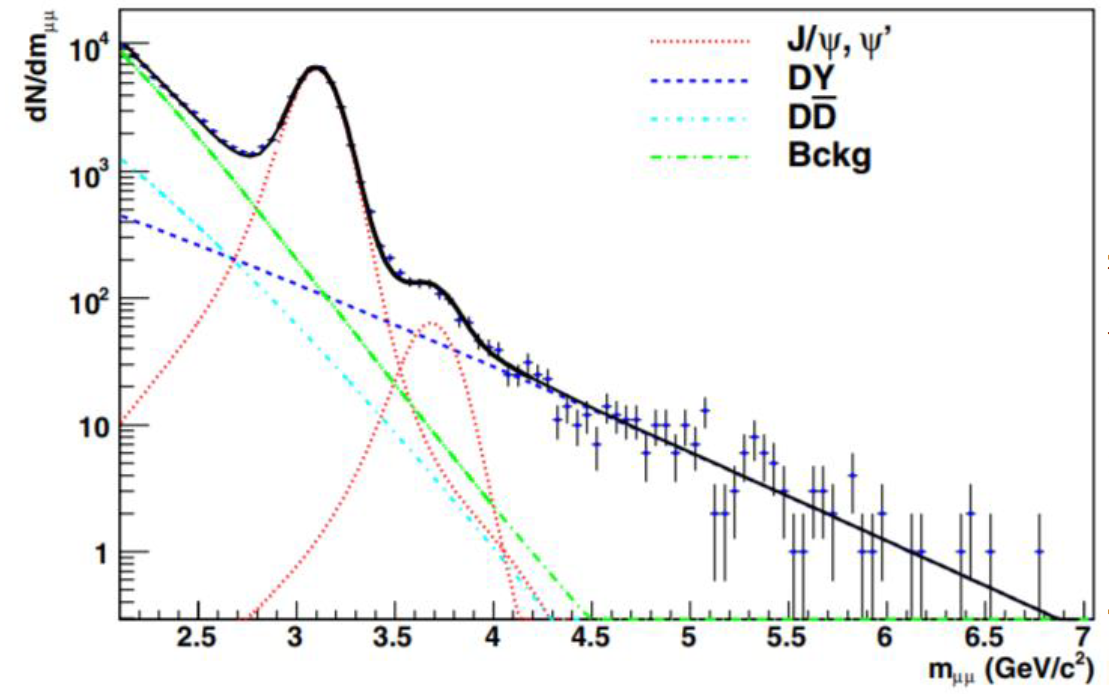}
	\caption{The invariant mass distribution of $ \mu^+\mu^- $ pairs produced in In+In collisions at 158\AGeVc showing a strong peak due to $ \text{J}/\psi $ decays \cite{PhysRevLett.99.132302}.}
	\label{fig:NA60Jpsi}
\end{figure}

Moreover, one could combine this measurements with published results on $ \text{J}/\psi $ production. The NA60 experiment \cite{PhysRevLett.99.132302} measured $ \text{J}/\psi $ production in In+In collisions (Fig.~\ref{fig:NA60Jpsi}), which is a system of similar size as Xe+La. This combination of the NA60 data on $ \text{J}/\psi $ and the \NASixtyOne results on open charm could already challenge theoretical models.

The SAVD is also planned to be used during three weeks of the Pb+Pb data taking in 2018 (recommended by the CERN SPS Committee in October 2017). About $1\cdot10^7$ central collisions should be recorded and 4000 \PDzero and \APDzero decays can be expected to be reconstructed in this data set. The expected signal to background ratio is about 2.6 with realistic PID.
\newpage
\section{Systematic studies of charm by \NASixtyOne} \label{sc:futureprogram}

Precise open charm measurements in \NASixtyOne are planned for the years 2021-2024. They will include measurements in central Pb+Pb collisions at 40\AGeVc and 150\AGeVc, in mid-central and peripheral Pb+Pb collisions at 150\AGeVc, and reconstruction of decays of various open charm mesons (see Table \ref{tab:opencharm}).

\subsection{\NASixtyOne upgrades}

\begin{figure}[!htb]
	\centering
	\includegraphics[width=0.8\textwidth]{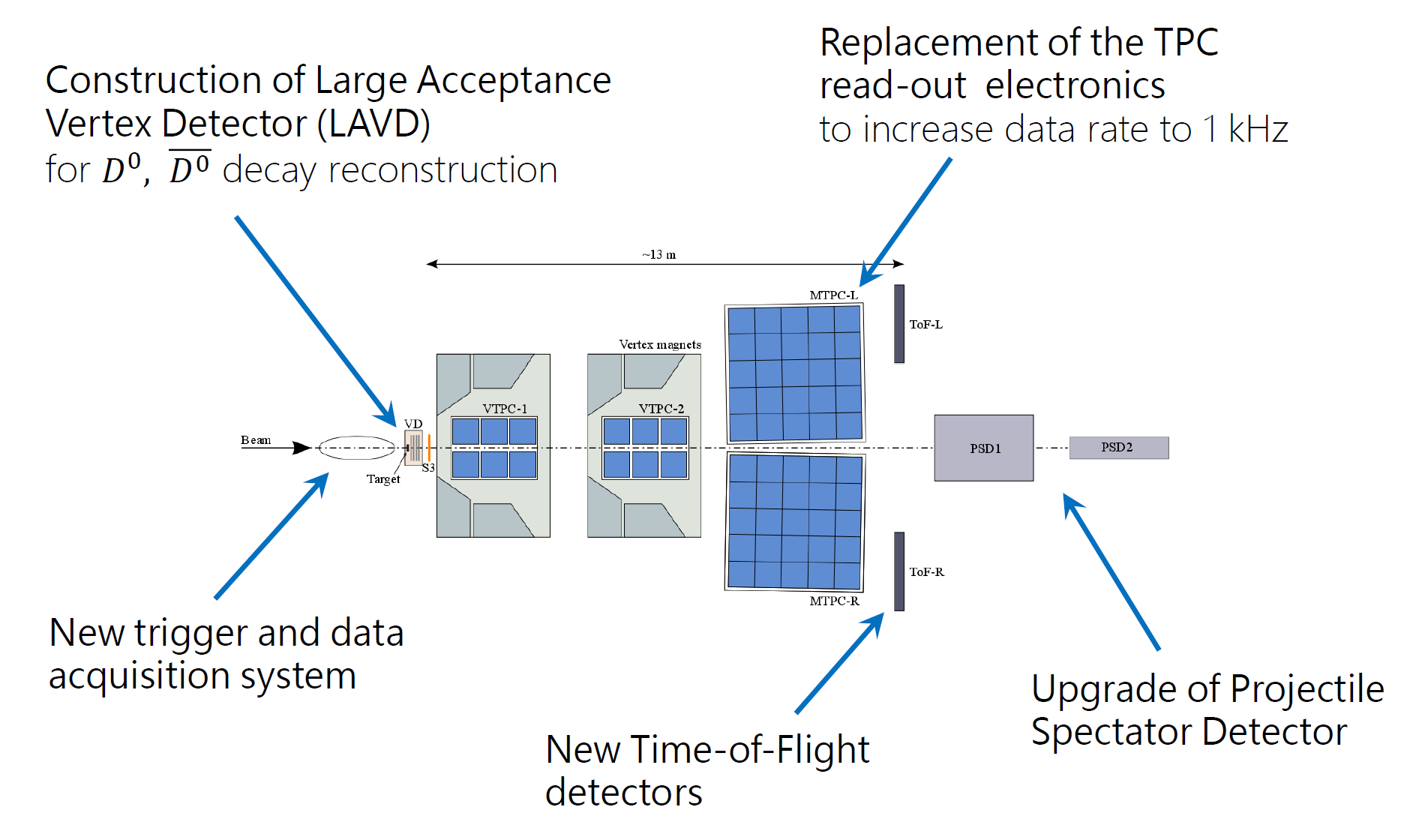}
	\caption{Upgrades of \NASixtyOne planned to be completed during the LS2 period.}
	\label{fig:NA61upgrade}
\end{figure}

During the Long Shutdown 2 at CERN (2019-2020), a significant upgrade of the \NASixtyOne spectrometer is planned (Fig. \ref{fig:NA61upgrade}), in order to allow precise charm measurements. 

The most important upgrade from the open charm point of view is the construction of a new vertex detector: the Large Acceptance Vertex Detector (LAVD). Also, for obtaining higher data rate (\SI{1}{\kilo\hertz}) the TPC electronics will be replaced and a new trigger and data acquisition system will be built. An upgrade of the Projectile Spectator Detector will also be performed to cope with the higher beam intensity expected to be delivered after LS2. Finally, new ToF detectors are planned to be constructed to improve particle identification at mid-rapidity.

\subsection{Large Acceptance Vertex Detector} \label{ssc:lavd}

To improve the quality of open charm measurement, the Large Acceptance Vertex Detector will replace the existing SAVD detector. The LAVD will be constructed using technology developed for the ALICE ITS: CMOS ALPIDE pixel sensors, carbon fiber support structures, and read-out electronics \cite{AglieriRinella:2017lym}. Basic properties of the ALPIDE sensors are: $ 28 \times \SI{28}{\micro\meter} $ pixels, less than $ \SI{10}{\micro\second} $ time resolution, $ 15 \times \SI{30}{\milli\meter^2} $ surface, 0.5 MPixel, and \SI{50}{\micro\meter} thickness. 

The estimated material budget per layer is $ 0.3\% $ of the radiation length. The detector will consists of 4 stations. The transverse dimensions of the stations are: $ 2 \times \SI{4}{\centi\meter^2} $ for the first station, $ 4 \times \SI{8}{\centi\meter^2} $ for the second, $ 6 \times \SI{12}{\centi\meter^2} $ for the third, and $ 8 \times \SI{16}{\centi\meter^2} $ for the fourth station. Each station has a square beam hole of ($ 3 \times \SI{3}{\milli\meter^2} $) in the center. The LAVD area will be covered by 126 sensors.

The beam request related to the open charm measurements in 2022-2024 is shown in Table \ref{tab:beamrequest}.

\renewcommand{\arraystretch}{1}

\begin{table}[!htb]
	\caption{The \NASixtyOne beam request related to the charm program.}
	\label{tab:beamrequest}
	\centering
	\begin{tabular}{l l l l l }
		Year & Beam & Duration &  Purpose & \PDzero + \APDzero stat. \\ \hline
		2021 & \Pproton at 150\AGeVc & 4 weeks & detector commissioning and tests & ~ \\
		2022 & Pb at 150\AGeVc & 2 weeks & charm in central collisions & 40k \\
		2022 & Pb at 150\AGeVc & 4 weeks & charm in peripheral collisions & 8k \\
		2023 & Pb at 150\AGeVc & 2 weeks & charm in mid-central collisions & 20k \\
		2024 & Pb at 40\AGeVc & 4 weeks & charm in central collisions & 2k \\
	\end{tabular}
\end{table}

\begin{figure}[!htb]
	\centering
	\includegraphics[width=0.55\textwidth]{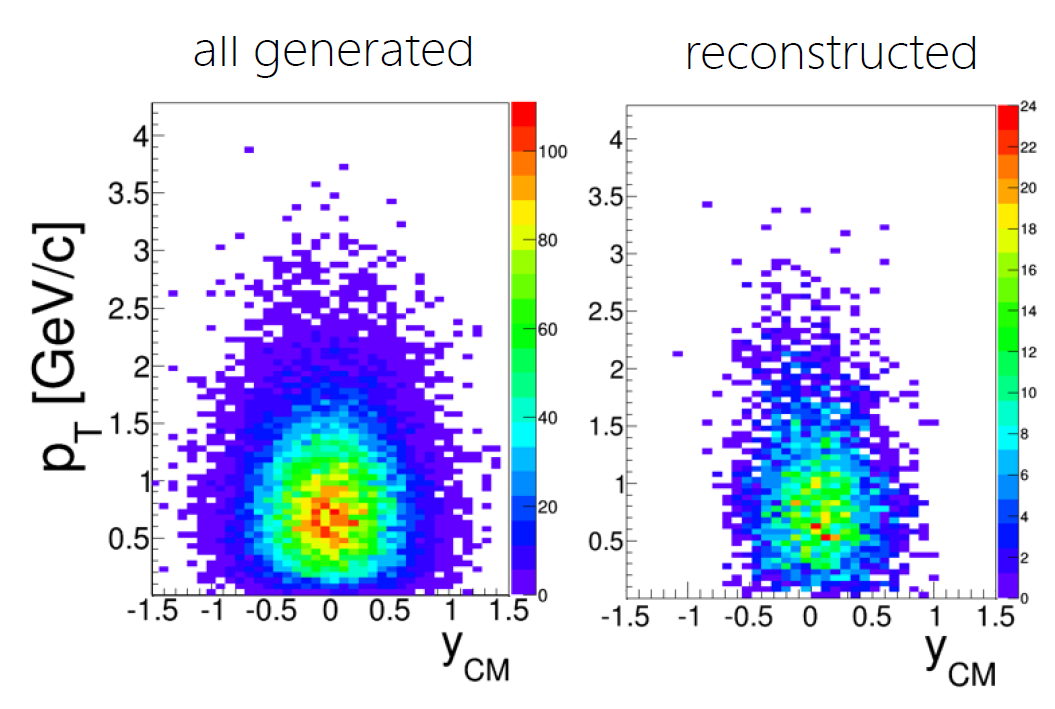}
	\caption{The AMPT simulation of transverse momentum and rapidity distributions of \PDzero mesons produced in central Pb+Pb collisions at 150\AGeVc corresponding to $4\cdot10^6$ events. \textit{Left}: all produced \PDzero mesons. \textit{Right}: \PDzero mesons fulfilling the following criteria: decay into $\PDzero \to \pi^+ + \PKminus$, decay products registered by the LAVD, and passing background suppression and quality cuts \cite{Aduszkiewicz:2059811}.}
	\label{fig:acceptSimLAVD}
\end{figure}

With the upgraded \NASixtyOne spectrometer, one expects that during two weeks of data taking in 2022, $4\cdot10^7$ central Pb+Pb collisions at 150\AGeVc will be collected and $4\cdot10^4$ \PDzero and \APDzero decays will be reconstructed.

The expected high statistics of reconstructed \PDzero and \APDzero decays is due to the high event rate and the large acceptance of the LAVD. Its acceptance will be about 12\% (three times better than for the SAVD) of all the \PDzero in the $\pi^+ + \PKminus$ decay channel. Moreover, based on AMPT simulations one estimates that fully corrected results will correspond to more than 90\% of the \PDzero and \APDzero yield (see Fig.~\ref{fig:acceptSimLAVD}). Total systematic uncertainty of $ \langle \PDzero \rangle $ and $ \langle \APDzero \rangle $ is expected to be about 10\%.

\subsection{Charm program: anticipated results} \label{ssc:charmresults}

The data obtained by \NASixtyOne should lead to unique and important results on charm production in heavy ion collisions.

\begin{figure}[!htb]
	\centering
	\includegraphics[width=0.65\textwidth]{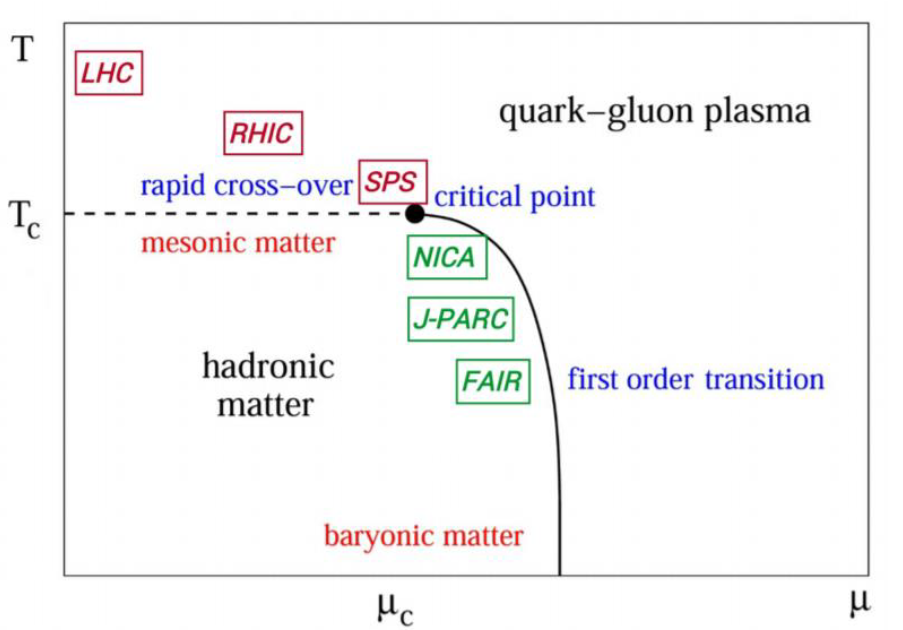}
	\caption{Present (red) and future (green) heavy ion facilities in the phase diagram of strongly interacting matter.}
	\label{fig:landscape}
\end{figure}

Figure \ref{fig:landscape} presents a compilation of present and future facilities and their region of coverage in the phase diagram of strongly interacting matter. Their capability to measure charm hadrons is summarized below:
\begin{enumerate}[label=(\roman*)]
	\item LHC and RHIC at high energies ($ \sqrt{s_{\text{NN}}} \gtrsim \SI{200}{\giga\electronvolt} $): measurements of open charm is performed in limited acceptance; this is due to the collider kinematics and related to the detector geometry \cite{Meninno:2017ezl, Hou:2016hgs, Simko:2017jtj, Nagashima:2017mpm}.
	\item RHIC BES collider ($ \sqrt{s_{\text{NN}}} =  7.7 - \SI{39}{\giga\electronvolt} $): measurement  not considered in the current program, this may be likely due to difficulties related to collider geometry and kinematics as well as the low charm production cross-section~\cite{Odyniec:2013kna, Yang:2017llt}.
	\item RHIC BES fixed-target ($\sqrt{s_{\text{NN}}} = 3 - \SI{7.7}{\giga\electronvolt} $): not considered in the current program~\cite{Meehan:2016qon}.
	\item NICA ($\sqrt{s_{\text{NN}}} < \SI{11}{\giga\electronvolt} $): measurements during the stage 2 (after 2023) are under consideration \cite{Kekelidze:2017tgp,KekelidzeWarsaw}.
	\item J-PARC-HI ($\sqrt{s_{\text{NN}}} \lesssim \SI{6}{\giga\electronvolt} $): under consideration, may be possible after 2025 \cite{Sako:2016edz,KitazawaReimei}.
	\item FAIR SIS-100 ($\sqrt{s_{\text{NN}}} \lesssim \SI{5}{\giga\electronvolt} $): not possible due to the very low cross-section at SIS-100, systematic charm measurements are planned with SIS-300 ($\sqrt{s_{\text{NN}}} \lesssim \SI{7}{\giga\electronvolt} $) which is agreed-on part of FAIR, but not of the start version (timeline is unclear) \cite{Friman:2011zz, VolkerTrento}. 
\end{enumerate}

The conclusion is that only NA61/SHINE is able to measure open charm production in heavy ion collisions in full phase space in the near future.\\

Having this data, one can try to answer the motivating questions.

\begin{figure}[!htb]
	\centering
	\includegraphics[width=0.95\textwidth]{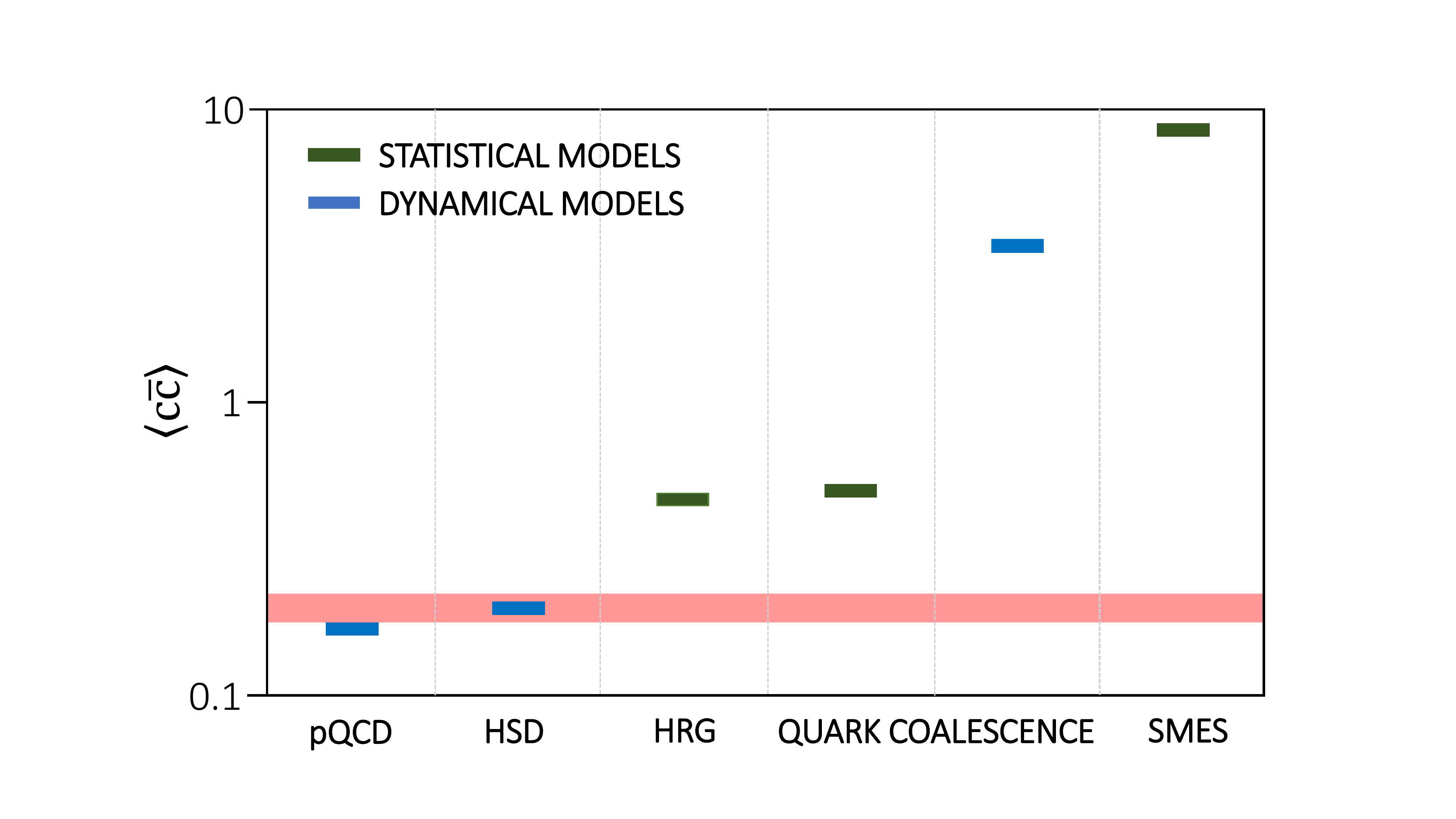}
	\caption{Mean multiplicity of charm quark pairs produced in central Pb+Pb collisions at 158\AGeVc calculated within dynamical models (blue bars) and statistical models (green bars). The width of the red band, at the location assuming HSD predictions, indicates the accuracy of the NA61 2020+ result.}
	\label{fig:answerModels}
\end{figure}
	
Figure \ref{fig:answerModels} shows a foreseen accuracy of \NASixtyOne data on mean charm multiplicity in reference to charm production models (see Sec. \ref{sc:motivation} for details). The red band indicates the foreseen accuracy of \NASixtyOne result on the charm yield assuming the yield prediction of the HSD model \cite{Linnyk:2008hp}. With that accuracy it should be possible to exclude most of the current models.

\begin{figure}[!htb]
	\centering
	\includegraphics[width=0.6\textwidth]{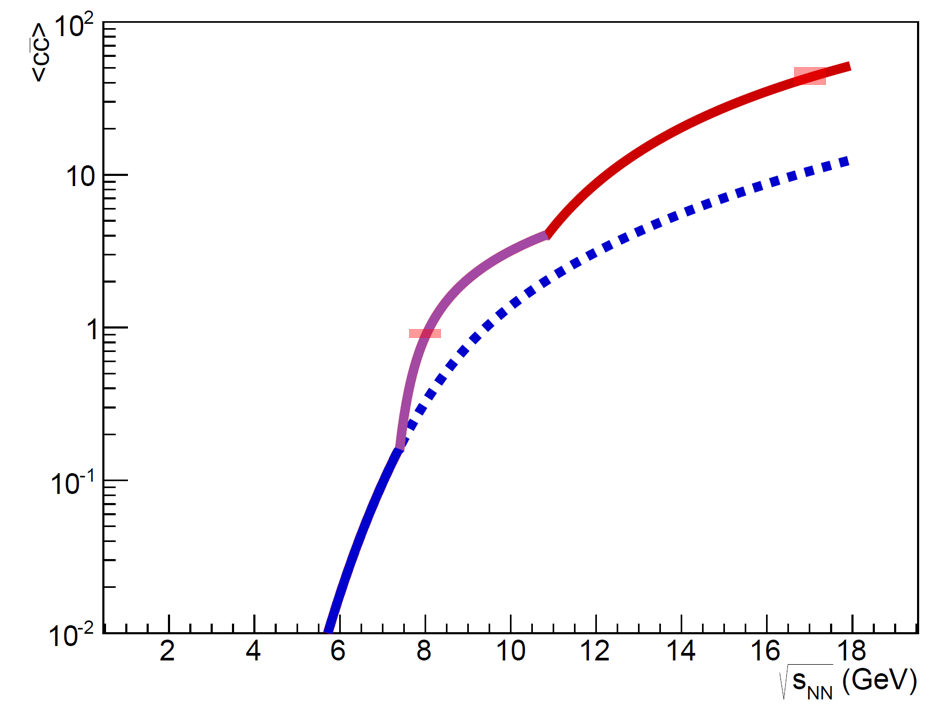}
	\caption{Energy dependence of $ \langle \Pcharm\APcharm \rangle $ in central Pb+Pb collisions calculated within the SMES model. The red bars indicate the accuracy of NA61 2020+ results for two energies: 40\AGeVc ($ \sqrt{s_{\text{NN}}} = 8.6~\GeV $) and 150\AGeVc ($ \sqrt{s_{\text{NN}}} = 16.7~\GeV $), assuming the SMES yields  \cite{Gazdzicki:1998vd}.}
	\label{fig:answerOnset}
\end{figure}

Figure \ref{fig:answerOnset} shows a foreseen accuracy of \NASixtyOne data on $ \langle \Pcharm\APcharm \rangle $ in reference to SMES predictions (see Sec. \ref{sc:motivation} for details). Red bars indicate the foreseen accuracy of the \NASixtyOne result on charm multiplicity for energies included in the proposed \NASixtyOne charm studies: 40\AGeVc ($ \sqrt{s_{\text{NN}}} = 8.6~\GeV $) and 150\AGeVc ($ \sqrt{s_{\text{NN}}} = 16.7~\GeV $) . These points would be a start of confronting model predictions on collision energy dependence, but measurements at more energies are necessary. In the future, this can be performed at J-PARC-HI and FAIR SIS-300.

\begin{figure}[!htb]
	\centering
	\includegraphics[width=0.6\textwidth]{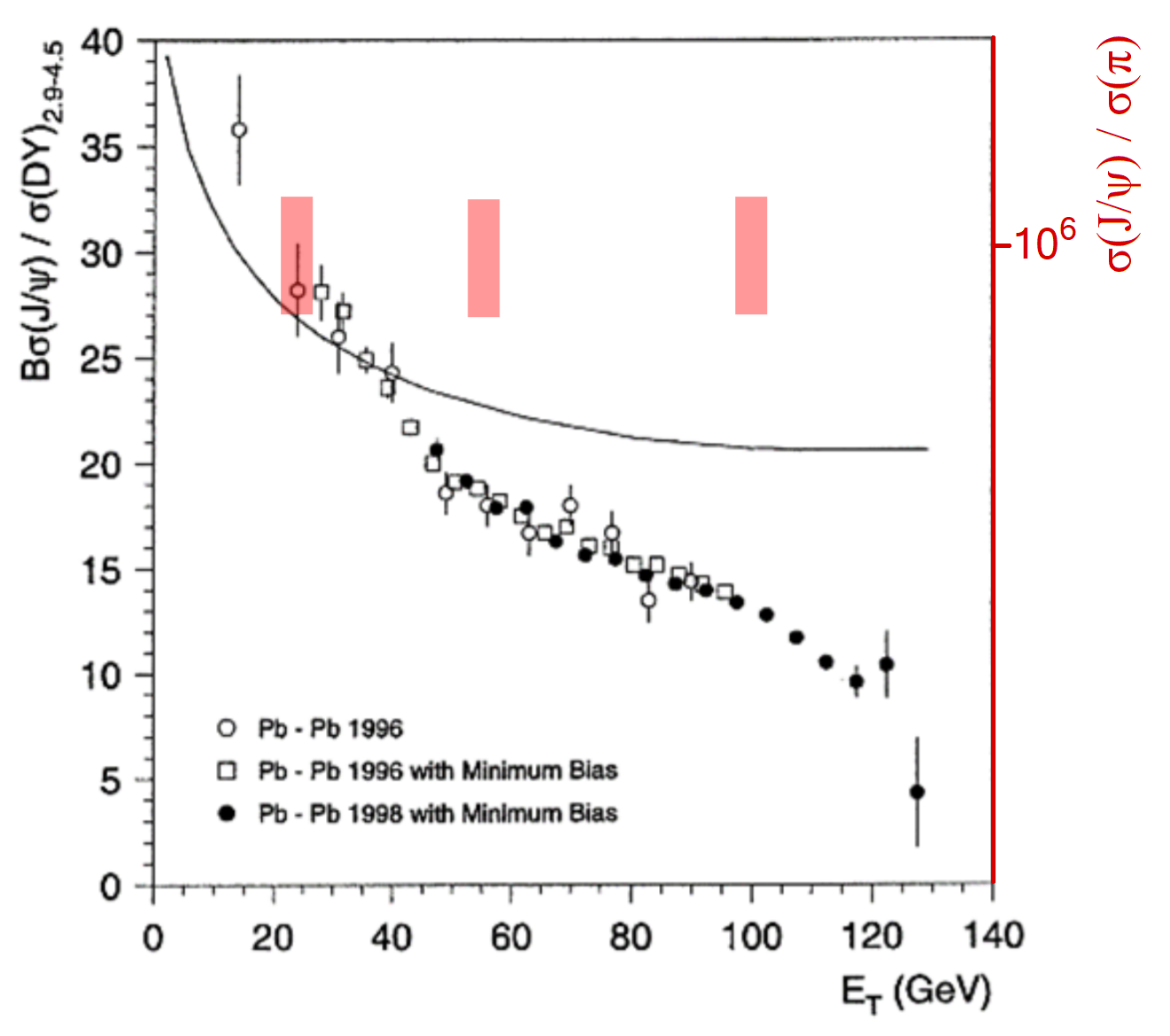}
	\caption{The ratio of $\sigma_{\text{J}/\psi}/\sigma_{\mathrm{DY}}$ (\textit{left}) and $\sigma_{\text{J}/\psi}/\sigma_{\pi}$ (\textit{right}) as a function of transverse energy in Pb+Pb collisions at 158\AGeV. The $\sigma_{\text{J}/\psi}/\sigma_{\mathrm{DY}}$ ratio was measured by NA50 \cite{Abreu:2000ni} and it was used to calculate the $\sigma_{\text{J}/\psi}/\sigma_{\pi}$ ratio in Ref. \cite{Gazdzicki:1998jx}. Red bars mark the $\sigma_{\text{J}/\psi}/\sigma_{\Pcharm \APcharm}$ accuracy of the NA61 2020+ result assuming $\sigma_{\Pcharm \APcharm} \sim \sigma_{\pi} $  and scaling to $\sigma_{\text{J}/\psi}/\sigma_{\mathrm{DY}}$ for peripheral collisions.}
	\label{fig:anwerJpsi}
\end{figure}

Figure \ref{fig:anwerJpsi} shows a foreseen accuracy of \NASixtyOne data on charm in reference to $ \text{J}/\psi $ suppression (see Sec. \ref{sc:motivation} for details). The red bars indicate the foreseen accuracy of the $\sigma_{\text{J}/\psi}/\sigma_{\Pcharm \APcharm}$ result that was made assuming $\sigma_{\Pcharm \APcharm} \sim \sigma_{\pi} $. One will be able to distinguish between two extreme scenarios: $ \langle \Pcharm\APcharm \rangle \sim \langle \mathrm{DY} \rangle $ or $ \langle \Pcharm\APcharm \rangle \sim \langle \pi \rangle $.

\newpage
\section{Summary} \label{sc:summary}

\NASixtyOne has started measurements of open charm production in 2017 with data taking on Xe+La collisions at 150\AGeVc. The measurements in central Pb+Pb collisions at 150\AGeVc will be performed in 2018. This has been possible due to an implementation of the silicon pixel Vertex Detector. During the Long Shutdown 2, \NASixtyOne plans to construct the Large Acceptance Vertex Detector and increase the data taking rate by a factor of 10, to \SI{1}{\kilo\hertz}. 

This will allow to collect high statistics data on open charm hadron production in central Pb+Pb collisions at 150\AGeVc and 40\AGeVc and centrality selected Pb+Pb collisions at 150\AGeVc. The results are expected to:
\begin{enumerate}[label=(\roman*)]
	\item distinguish between many existing models of open charm production in Pb+Pb collisions,
	\item initiate a measurement of collision energy dependence of open charm yield, which may lead to an observation of open charm signal of the onset of deconfinement, 
	\item verify $ \text{J}/\psi $ signal of the QGP formation by measurements of centrality dependence of charm production.
\end{enumerate}

\newpage
\section*{Acknowledgments}

We would like to thank Elena Bratkovskaya, Giuseppe Bruno, Mark Gorenstein, Paolo Martinengo, Roman Poberezhnyuk, and Taesoo Song for the information, comments and support that helped to prepare this document.
\\

This work was supported by the Hungarian Scientific Research Fund (Grants
NKFIH 123842--123959), the J\'anos Bolyai Research Scholarship
of the Hungarian Academy of Sciences, the Polish Ministry of Science
and Higher Education (grants 667\slash N-CERN\slash2010\slash0,
NN\,202\,48\,4339 and NN\,202\,23\,1837), the Polish National Center
for Science (grants~2011\slash03\slash N\slash ST2\slash03691,
2013\slash11\slash N\slash ST2\slash03879, 2014\slash13\slash N\slash
ST2\slash02565, 2014\slash14\slash E\slash ST2\slash00018,
2014\slash15\slash B\slash ST2\slash02537 and
2015\slash18\slash M\slash ST2\slash00125, 2015\slash 19\slash N\slash ST2 \slash01689), the Foundation for Polish
Science --- MPD program, co-financed by the European Union within the
European Regional Development Fund, the Federal Agency of Education of
the Ministry of Education and Science of the Russian Federation (SPbSU
research grant 11.38.242.2015), the Russian Academy of Science and the
Russian Foundation for Basic Research (grants 08-02-00018, 09-02-00664
and 12-02-91503-CERN), the Ministry of Science and
Education of the Russian Federation, grant No.\ 3.3380.2017\slash4.6,
 the National Research Nuclear
University MEPhI in the framework of the Russian Academic Excellence
Project (contract No.\ 02.a03.21.0005, 27.08.2013),
the Ministry of Education, Culture, Sports,
Science and Tech\-no\-lo\-gy, Japan, Grant-in-Aid for Sci\-en\-ti\-fic
Research (grants 18071005, 19034011, 19740162, 20740160 and 20039012),
the German Research Foundation (grant GA\,1480/2-2), the EU-funded
Marie Curie Outgoing Fellowship, Grant PIOF-GA-2013-624803, the
Bulgarian Nuclear Regulatory Agency and the Joint Institute for
Nuclear Research, Dubna (bilateral contract No. 4418-1-15\slash 17),
Bulgarian National Science Fund (grant DN08/11), Ministry of Education
and Science of the Republic of Serbia (grant OI171002), Swiss
Nationalfonds Foundation (grant 200020\-117913/1), ETH Research Grant
TH-01\,07-3 and the U.S.\ Department of Energy.

\newpage

\bibliographystyle{na61Utphys}
{\footnotesize\raggedright
	\bibliography{na61References}
}

\end{document}